\documentclass[
	aps, aps,pra,longbibliography, superscriptaddress, twocolumn,
	10pt
	floatfix, 
    nofootinbib,
	tightenlines
]{revtex4-1}
\usepackage[final]{graphicx}
\usepackage{times,bbm,amsmath,amssymb}
\usepackage{epsfig,color}
\usepackage{xcolor}
\usepackage{hyperref}
\hypersetup{
    colorlinks = true
}
\usepackage{cleveref}
\usepackage{microtype}

\usepackage{float,siunitx}
\usepackage[caption = false]{subfig}

\usepackage[greek,english]{babel}
\usepackage{thumbpdf,enumerate}
\usepackage{booktabs}
\usepackage{sidecap}
\usepackage[scaled=.8]{couriers}
\usepackage{multirow}
\usepackage{placeins}
\usepackage{relsize}
\usepackage{pst-grad,bm}
\usepackage{epigraph}
\usepackage{gensymb}
\usepackage{longtable}
\usepackage{ulem} 
\usepackage{acronym}
\usepackage{physics}
\usepackage{easyReview}
\usepackage[columnwise]{lineno}

\DeclareUnicodeCharacter{0301}{\'{e}}

\begin{document}

%\linenumbers
%\setlength\linenumbersep{0.2cm}$$

\vspace{10pt}

\title{Contrast enhanced imaging through weakly scattering media with spatially entangled photons}

\author{Alessio D'Errico} 
\email{aderrico@uottawa.ca}
\address{Nexus for Quantum Technologies, University of Ottawa, Ottawa, K1N 6N5, ON, Canada}
\affiliation{National Research Council of Canada, 100 Sussex Drive, K1A 0R6, Ottawa, ON, Canada}

\author{James Hubble$^\dagger$} 
\address{Nexus for Quantum Technologies, University of Ottawa, Ottawa, K1N 6N5, ON, Canada}
\affiliation{National Research Council of Canada, 100 Sussex Drive, K1A 0R6, Ottawa, ON, Canada}

\author{Rojan Abolhassani$^\dagger$} 
\address{Nexus for Quantum Technologies, University of Ottawa, Ottawa, K1N 6N5, ON, Canada}
\affiliation{National Research Council of Canada, 100 Sussex Drive, K1A 0R6, Ottawa, ON, Canada}

\author{Nazanin Dehghan}
\address{Nexus for Quantum Technologies, University of Ottawa, Ottawa, K1N 6N5, ON, Canada}
\affiliation{National Research Council of Canada, 100 Sussex Drive, K1A 0R6, Ottawa, ON, Canada}

\author{Yishai Klein} 
\address{Nexus for Quantum Technologies, University of Ottawa, Ottawa, K1N 6N5, ON, Canada}
\affiliation{National Research Council of Canada, 100 Sussex Drive, K1A 0R6, Ottawa, ON, Canada}

\author{Yingwen Zhang} 
\address{Nexus for Quantum Technologies, University of Ottawa, Ottawa, K1N 6N5, ON, Canada}
\affiliation{National Research Council of Canada, 100 Sussex Drive, K1A 0R6, Ottawa, ON, Canada}

\author{Ebrahim Karimi}
\address{Nexus for Quantum Technologies, University of Ottawa, Ottawa, K1N 6N5, ON, Canada}
\affiliation{National Research Council of Canada, 100 Sussex Drive, K1A 0R6, Ottawa, ON, Canada}
\affiliation{Institute for Quantum Studies, Chapman University, Orange, California 92866, USA}

\begin{abstract}
Improving the image contrast of objects immersed in weakly scattering media can be achieved using various strategies. One common approach is to reject events associated with scattered photons in favor of the detection of ballistic photons. While this is traditionally done via time gating or spatial filtering, we propose a different approach based on probing the object with spatio-temporally entangled photon pairs. We show that coincidence detection, followed by post-selection on spatially correlated events, allows us to isolate ballistic from scattered bi-photons, thereby enhancing image contrast relative to a single-photon detection strategy, and simultaneously removes events due to background light. Our predictions are obtained via numerical simulations and confirmed by experiments conducted in two configurations where either both photons or only one illuminates the scene. In both scenarios, correlation post-selection shows an improvement in image contrast at the expense of higher shot noise due to the lower number of events. The latter can be partially compensated for by appropriately combining events from several post-selection windows. Our findings will enable extending imaging through scattering media into the quantum imaging framework in settings where adaptive optics, time gating, and spatial filtering are impractical. 
\end{abstract}

\maketitle 
\def\thefootnote{$\dagger$}\footnotetext{These authors contributed equally to this work.}\def\thefootnote{\arabic{footnote}}

\section{Introduction}
Imaging techniques, from microscopy and light detection and ranging to astronomy, can be affected by unknown inhomogeneities in the medium between the source and detector, resulting in a loss of image quality, quantified in terms of resolution and/or contrast~\cite{rotter2017light, bertolotti2022imaging}. Whether the medium is a turbulent atmosphere, fog, a multi-mode waveguide, or tissue, the propagation of light is altered by phase aberrations, scattering, and coherence loss \cite{bertolotti2022imaging, akkermans2007mesoscopic,yoon2020deep}.

A common challenge in classical and quantum optical imaging is to devise techniques to address these undesired effects and recover image quality. %A broad range of approaches has been proposed, which can be listed in a few  categories: numerical analysis \cite{gu2015microscopic, faccio2020non,yoon2020deep}, where the knowledge of physical properties of the scattering medium can be employed to recover effective point spread functions,  correction methods \cite{leith1966holographic, popoff2010measuring, mounaix2016deterministic,horstmeyer2015guidestar, cao2022shaping}, where the scattering medium is experimentally characterized and its action is compensated for by means of spatial light modulators, and filtering techniques \cite{huang1991optical, schilders1998microscopic, dunsby2003techniques,pawley2006handbook, zipfel2003nonlinear, gu1997monte, gu2015microscopic}, when one tries to isolate the contributions in the light field that carry residual information about the object. 
A broad range of approaches has been proposed, which can be grouped into a few categories: filtering techniques \cite{huang1991optical, schilders1998microscopic, dunsby2003techniques, pawley2006handbook, zipfel2003nonlinear, gu1997monte, gu2015microscopic}, in which one aims to isolate the components of the light field that carry residual information about the object; numerical analysis \cite{gu2015microscopic, faccio2020non, yoon2020deep}, where knowledge of the physical properties of the scattering medium is used to recover effective point spread functions; and correction methods \cite{leith1966holographic, popoff2010measuring, mounaix2016deterministic, horstmeyer2015guidestar, cao2022shaping}, where the scattering medium is experimentally characterized and its effect is compensated for using spatial light modulators.

The extension of these approaches to the few-photon regime is demanded by the growing potential of quantum imaging techniques, which use entangled two-photon states and spatially resolved coincidence detection \cite{moreau2019imaging, defienne2024advances,derrico2026}. Two photon correlations in different degrees of freedom of light have been investigated in various complex media, revealing migration of entanglement in specific sets of scattering modes \cite{lib2022quantum,valencia2020unscrambling, courme2023quantifying}, unique features in the correlation speckles \cite{beenakker2009two,di2012statistical, klein2016speckle}, survival of polarization entanglement \cite{shi2016photon}, and enhanced effects, for instance, in coherent back-scattering \cite{safadi2023coherent}. Moreover, spatio-temporal correlations can be employed to devise novel adaptive optics techniques directly applicable at the single photon level \cite{defienne2014nonclassical,defienne2018adaptive, shekel2021shaping,cameron2024adaptive,mccarthy2009long,bajar2025partial}.

\begin{figure}[!t]
\includegraphics[width=0.8\columnwidth]{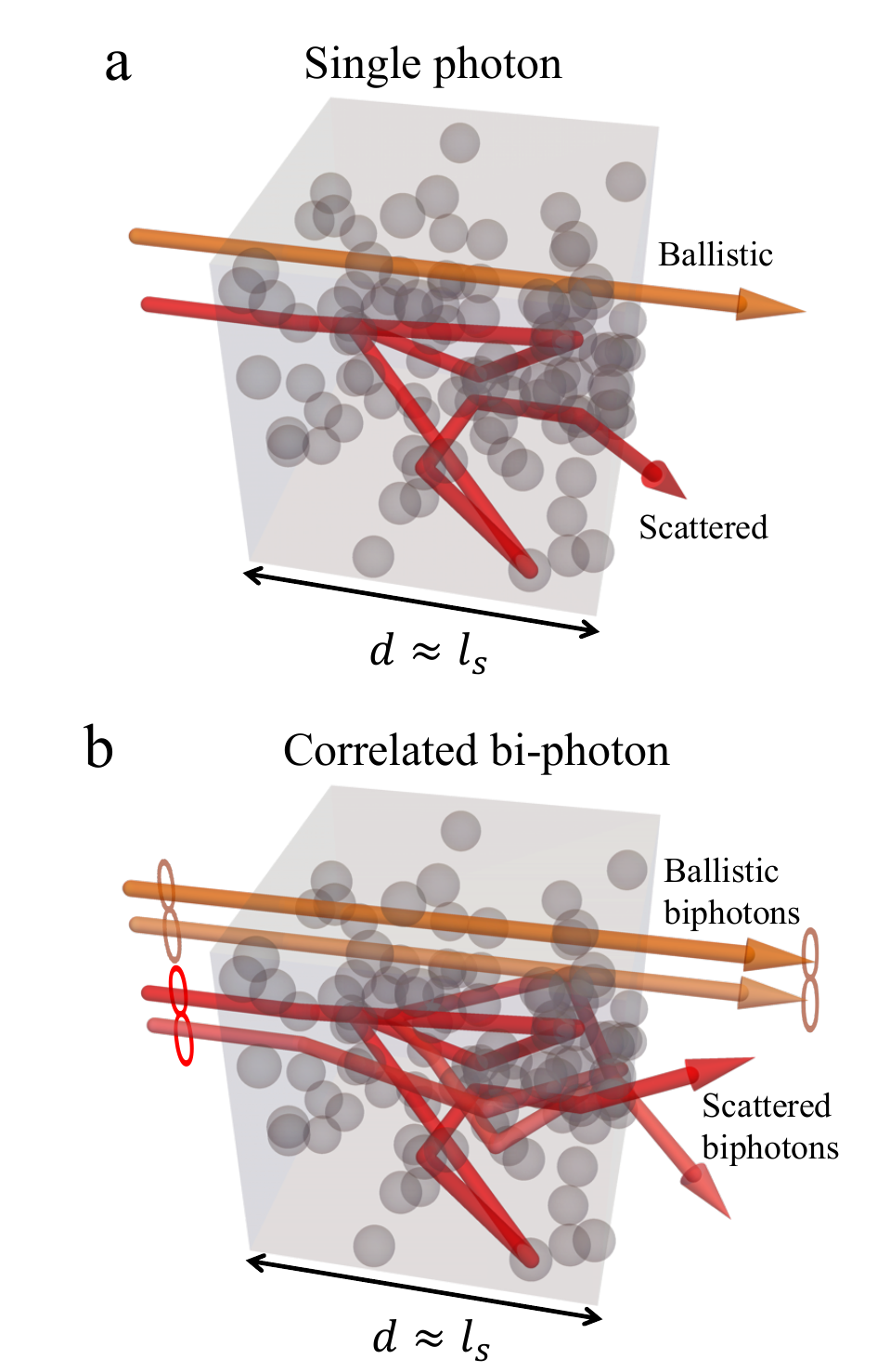}
\caption{{\bf Conceptual scheme}. {\bf a,} In weakly scattering media (thickness $d$ no longer than $\sim 5$ times the mean free path $l_s$) is possible to identify ballistic photons and isolate them from scattering events either using time gating or spatial filtering. Ballistic photons carry information on objects hidden behind or inside the scattering medium. {\bf b,} In this work, we propose exploiting the fact that ballistic photon pairs preserve transverse position entanglement, and, thus, can be isolated from scattered biphotons via correlation post-selection.} 
\label{fig:idea}
\end{figure}

Hitherto, the introduction of filtering techniques in the quantum optical regime has been mostly limited to time-correlation-based methods \cite{ou2007multi}, specifically in quantum LIDAR experiments \cite{mccarthy2009long}. In classical optics, filtering techniques try to distinguish and isolate different light paths through the medium. Generally speaking, one can distinguish between ballistic photons, which propagate undisturbed through the medium, and scattered photons, whose optical paths are modified one or more times by scattering centers (Figure \ref{fig:idea}-a). In \cite{gu1997monte}, the authors also define ``snake-photons'' corresponding to photons whose optical path is only weakly modified with respect to the ballistic photons. Filtering methods are effective in weakly scattering regimes, where the thickness of the medium is at most of the same order as the mean scattering length, $l_s$, so that the ballistic photon contributions are still significant. When using pulsed light illumination, scattered photons can be filtered out using time-gating techniques \cite{duguay1969ultrafast,hee1993femtosecond,xu2026epsilon}, which exploit the fact that ballistic photons are the first to reach the detector. Under continuous-wave (cw) illumination, spatial or coherence filtering techniques can also be used \cite{huang1991optical}. Furthermore, these techniques can be combined with polarization filtering, where one makes use of the fact that scattered photons can accumulate a different polarization with respect to the input radiation \cite{gan1999image}. 

%%%%%%%%%%%%%%%%%%%%%
%%%%%%%%%%%%%%%%%%%%%
%In the case of thin scattering media the main role of scattering is to introduce transversely dependent random phase shifts. In this regime, time gating requires the use of ultrashort pulses or very low coherence length, polarization filtering would be ineffective in absence of relevant anisotropies, and spatial filtering will lead to a loss in resolution. at the same time a thin scattering layer can introduce important losses in image contrast, and thus alternative methods are needed to isolate ballistic from scattered light. 

In thin scattering media, scattering primarily manifests as transversely varying random phase shifts. In this regime, conventional rejection strategies are often ineffective or impractical: time gating requires ultrashort pulses or broadband light with very low coherence length, polarization filtering provides little benefit in the absence of significant anisotropy, and spatial filtering inevitably sacrifices resolution. At the same time, even a thin scattering layer can substantially reduce image contrast, making it necessary to develop alternative methods for discriminating ballistic light from scattered contributions.
%In this work we address this issue in a quantum imaging scenario where spatially correlated photon pairs are used as the light source. Our approach, as illustrated in Fig. \ref{fig:idea}-b, relies on the fact that, when spatially correlated photon pairs propagate through a scattering medium, the correlations are broadened or lost. However, photon pairs undergoing a ballistic or snake path will preserve their spatial correlations. Our strategy is thus to filter out scattered biphotons by selecting coincidence events where the transverse coordinates of the two photons are the same as before the propagation through the medium. For instance, if the two photons originate from the same point, then we select coincidence events where the two photons are detected in the same transverse position. Our approach, thus, combines quantum spatial correlations and postelection techniques yielding a contrast enhancement of images of objects immersed in a weakly scattering medium.

Here, we address this problem using spatially entangled photon pairs as the illumination source. As illustrated in Fig.~\ref{fig:idea}-b, our approach exploits the fact that propagation through a scattering medium degrades the spatial correlations between entangled photon pairs. By contrast, photon pairs that follow ballistic or near-ballistic, ``snake''-like paths can largely preserve their initial spatial correlations. We, therefore, suppress scattered biphoton contributions by postselecting coincidence events whose detected transverse coordinates remain consistent with the correlations before propagation through the medium. In this way, our method combines quantum spatial correlations with coincidence-based postselection to enhance the imaging contrast of objects embedded in weakly scattering media.
%We first verify our predictions with numerical simulations, where we investigate how the contrast enhancement is affected by the scattering strength and the source entanglement. Then we experimentally benchmark the results by demonstrating the contrast enhancement under both dynamic and static scattering configurations. 

We first verify our predictions through numerical simulations, examining how the achievable contrast enhancement depends on the scattering strength and the degree of source entanglement. Then, we experimentally benchmark the method by demonstrating contrast enhancement under both dynamic and static scattering conditions, using two configurations: one in which both photons illuminate the object and propagate through the scattering medium, and another in which only one photon of the pair interacts with the object and undergoes scattering while the other acts as a reference.

%%%%%%%%%%%%%%%%%
\begin{figure*}[t]
\includegraphics[width=0.8\textwidth]{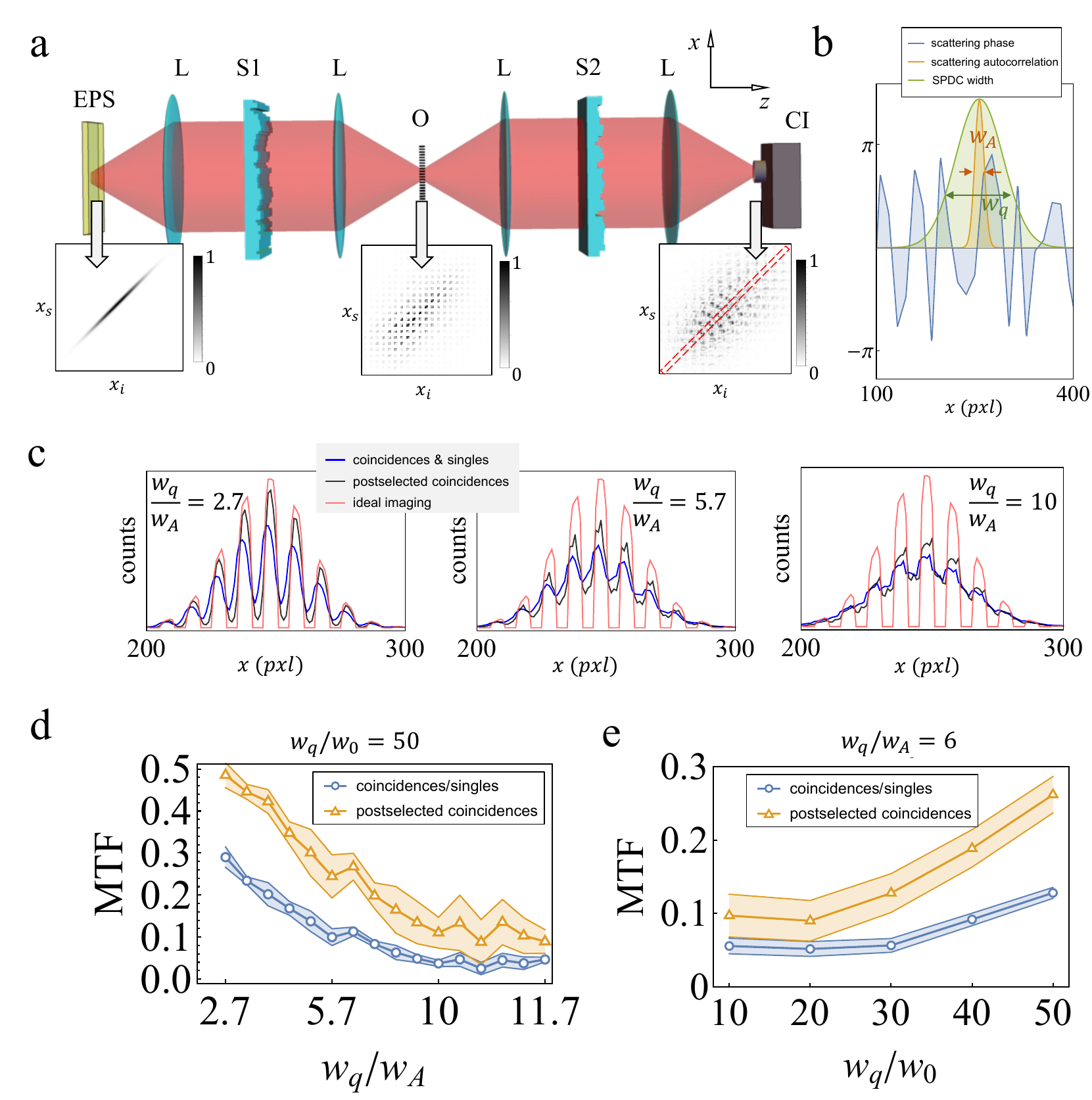}
\caption{{\bf Numerical simulations.} {\bf a,} Scheme of the simulated setup: spatially entangled photon pairs (EPS) are propagated in free-space to a random phase screen (S1) placed in the far field of the source. Then the pair is incident on an object (O), with spatially varying transmissivity, placed in the source image plane. A second random phase screen (S2) is placed in the object's far field. Coincidence imaging (CI) is performed on the source/object image plane. L indicates lenses performing optical Fourier transforms. Insets show the spatial correlations (for a single scattering realization) along the $x$-coordinate in different propagation planes. The red dashed rectangle indicates the numerically post-selected region in the correlation space. {\bf b,} Plots illustrating the definitions and relative size comparison of characteristic widths of the problem. The far-field anti-correlation width $w_q$ is compared with the autocorrelation width $w_A$ that characterizes the scattering strength of the phase masks considered (the blue line shows a single realization of the scattering). {\bf c,} Reconstructed images of the object for different relative scattering strengths obtained after averaging the spatial correlations over 100 scattering realizations. Red curves indicate the ground truth that would be observed in the absence of scattering. Blue curves are obtained by summing the correlation plots along the $x_i$ coordinate; this curve is the same for the cases of using singles or total coincidence counts. Black curves instead are obtained from spatial correlation post selection, as shown by the diagonal region outlined by the red dashed line in panel {\bf a}. {\bf d,} Image contrast (measured by the MTF in Eq. \eqref{eq:mtf}) as a function of the relative scattering strength $w_q/w_A$ for fixed input correlation strength $w_q/w_0=50$. The contrast is higher for post-selected coincidences at all considered strengths, although the advantage reduces for strong scattering, where ballistic biphotons become scarce. Panel {\bf e} shows that, for a fixed relative scattering strength, the contrast enhancement grows monotonically with the strength of the spatial correlations of the input state. Results in d and e are averaged over multiple simulation runs. Each run considered 100 scattering realizations. Colored bands show the standard deviations over all runs. All counts are normalized and reported in arbitrary units.  } 
\label{fig:sim}
\end{figure*}
%%%%%%%%%%%%%

\section{Simulations}
We consider the simplified problem of imaging an object O, characterized by a spatially variable transmission function $O(x)$, immersed in a weakly scattering medium consisting of random phase masks (S1 and S2) placed in the far field of O. The masks are placed both before and after the object, as illustrated in Figure \ref{fig:sim}-a. The choice to place S1 and S2 in planes conjugate to the object plane is motivated by the fact that these phase distortions will be the ones most severely affecting the imaging system. The system is probed by a pair of frequency-degenerate, spatially entangled photons (EPS). Indicating with $q_i$, $q_s$ the transverse wave-vector components of idler and signal photons, the input state can be described by the bi-photon wavefunction
\begin{align}
    \psi(q_i,q_s)=\exp\biggl(-\frac{(q_i-q_s)^2}{w_q^2}\biggr) \exp\biggl(-\frac{(q_i+q_s)^2}{w_0^2}\biggr),
\end{align}
where we assumed a typical double-Gaussian approximation \cite{walborn2010spatial, derrico2026}, with $w_q$ and $w_0$ respectively indicating the characteristic width of the phase-matching function and the pump amplitude in momentum space. For simplicity, we consider a 1D spatial coordinate, this treating the two-photon propagation as a 2D problem. This significantly reduces the computational time for the simulations. 

In typical SPDC sources, $w_q\gg w_0$, which implies that photon pairs are strongly anti-correlated in transverse momentum and correlated in transverse position (considered in any plane where the source is imaged) \cite{walborn2010spatial, derrico2026}. The correlation strength can thus be measured from the ratio $w_q/w_0$. The random phase masks act directly in the transverse momentum space -- which, apart from a scaling factor, correspond to the far-field of the source -- as follows:
\begin{align}
\psi(q_i,q_s)\rightarrow e^{i(\varphi^{(r)}(q_i)+\varphi^{(r)}(q_s))}\psi(q_i,q_s),
\end{align}
where $\varphi^{(r)}(q)$ is a random phase, with the superscript $r=1,\ldots, R$ indicating a specific realization selected from an ensemble of $R$ elements. In both simulations and experiments, these phase screens were generated as lists of uniformly distributed random numbers in the interval $[-\pi,\pi]$. The number of elements in the list approximately fixed the scattering strength, while, prior to the simulations, the list was interpolated to extract an array with the same resolution as the chosen simulation window $M=2^9$ pixels -- see Supplementary Figure S2 for a detailed illustration. 

To quantify the scattering strength, we first calculate the disorder-averaged autocorrelation function of the phase masks $A(x)=\frac{1}{R}\sum_r\sum_{x'}\varphi^{(r)}(x-x')\varphi^{(r)}(x')$. For a sufficiently large $R$, $A(x)$ is well approximated by a Gaussian of width $w_A$. 
This allows us to quantify the scattering strength by the ratio $w_q/w_A$: If $w_q<w_A$, the two photons, despite being spatially anti-correlated, will, on average, experience similar phase shifts and will still be strongly correlated in the near-field. Instead, for $w_q>w_A$ (see Fig. \ref{fig:sim}-b), each photon will acquire a randomly different phase factor. The Fourier transform of the resulting bi-photon wavefunction (corresponding to the state in the object plane) will be largely uncorrelated (see, e.g. inset in Fig. \ref{fig:sim}-a in correspondence of O). We note that the ratio $w_q/w_A$ is a convenient definition of the scattering strength since it can be directly estimated from correlation measurements (see Supplementary S1 and Supplementary Figure S5).

The detrimental role of the scattering before and after the object can be understood as follows: 1) The random mask S1 between source and object scrambles the two-photon correlations, hence affecting the advantage offered by sub-shot noise quantum imaging; 2) the mask S2 between object and detector reduces the image contrast also when incoherent illumination is employed.
For incoherent light, only S2 affects image quality, whereas for coincidence imaging of correlated photons, both S1 and S2 have undesired effects. Specifically, one could expect that, with S1 in place, we would lose any advantage in using spatially correlated photons. As we explain below, our numerical simulations show that this is not the case, and that the analysis of coincidence images of spatially entangled photons can yield contrast enhancement beyond what is allowed by standard intensity imaging.

The crucial idea is to collect coincidences $\Gamma(x_i,x_s)\propto\abs{\psi(x_i,x_s)}^2$ in the source/object image plane and select those events for which $x_i\approx x_s$. A \textit{coincidence image} is typically obtained by observing the marginal with respect to the position of the idler photon $\Gamma_s(x_s)=\sum_{x_i}\Gamma(x_i,x_s)$ (corresponding to the blue lines in Fig. \ref{fig:sim}-c). In the presence of scattering, the image contrast in $\Gamma_s$ will be strongly reduced. In this case, most scattered photon pairs will populate events outside the diagonal of the correlation matrix, while ballistic biphotons will contribute to events along the diagonal. We thus expect that the \textit{postselected image} $\Gamma_{post}:=\Gamma(x_i=x_s)$ contains a higher relative number of ballistic contributions compared to $\Gamma_s(x)$ and should thereby exhibit higher contrast. 

Figure \ref{fig:sim}-c shows some examples of this contrast enhancement where we simulate, for different scattering strengths, the imaging of an object consisting of a regular array of apertures. In all these cases, the postselected coincidences $\Gamma_{post}(x)$ exhibit a higher contrast with respect to the coincidence image $\Gamma_s(x)$ (which, in the context of these simulations, is also formally equivalent to the expected result of intensity measurements). All the results shown are obtained from coincidences averaged over multiple scattering realizations ($R=100$), which corresponds to experimentally imaging objects immersed in dynamic scattering media -- the static case is illustrated in Supplementary figure S1. 

To estimate the contrast enhancement more quantitatively, we measure the resulting contrast through a Modulation Transfer Function (MTF) defined as follows. Let $I_{scatt}(x)$ be the coincidence image with the scattering medium and $I_{true}(x)$ the ground truth, i.e. the image with no scattering. The object chosen for the simulations is such that the Fourier transform of $I_{true}(x)$ has one dominant non-zero spatial frequency $k_0$. $I_{scatt}(x)$ will have lower frequency contributions due to the reduced image contrast. We define the MTF as the ratio between the nonzero and the zero frequency contributions: 
\begin{align}\label{eq:mtf}
\text{MTF}=\frac{\max\{|FT[I_{scatt}](k\approx k_0)|\}}{\max\{|FT[I_{scatt}](k\approx 0)|\}},   
\end{align}
with $FT$ indicating the Fourier transform. We thus evaluate the MTF for different scattering strengths (with a fixed correlation strength). Figure \ref{fig:sim}-d shows the MTF averaged over multiple runs (with the standard deviation given by the colored bands). It is evident that the post-selection correlation yields contrast enhancement for $w_q/w_A\lesssim 10$. For stronger scattering, the enhancement decreases due to the reduced number of ballistic bi-photons. In Fig. \ref{fig:sim}-e, we confirm that the contrast enhancement is allowed by the entanglement of the input two-photon state: stronger correlations (proportional to $w_q/w_0$) yield higher MTF enhancement when fixing the scattering strength. Note that, while we kept the ratio $w_q/w_0$ between 10 and 50 to avoid the necessity of high-resolution simulations, experimentally it is easy to reach ratios $w_q/w_0>100$ \cite{devaux2020imaging,gao2022high}. 

The simulations have been carried out considering only one transverse coordinate to bypass the computational difficulty of propagating the biphoton wavefunction dependent on the 4-spatial coordinates of idler and signal photons. We address the general case of two-dimensional objects and scattering phase-masks in the following experimental section.

%Second, correlation post-selection comes at the price of discarding many events which, especially in quantum imaging applications, can severely affect the acquisition time and result in relatively high shot noise of the post-selected images. This problem can be partially addressed by also extracting the correlation images $\Gamma_{post}(x;\,\xi):=\Gamma(x_i=x_s+\xi)$ with a variable shift $\xi$ which are mostly given by scattering events where one of the two photons was affected by a fixed tip-tilt aberration. These aberration only introduce a transverse shift in the object image without disrupting the contrast and can be deterministically corrected (see Experimental Results). 

\begin{figure*}[t!]
  \centering
  \includegraphics[width=2\columnwidth]{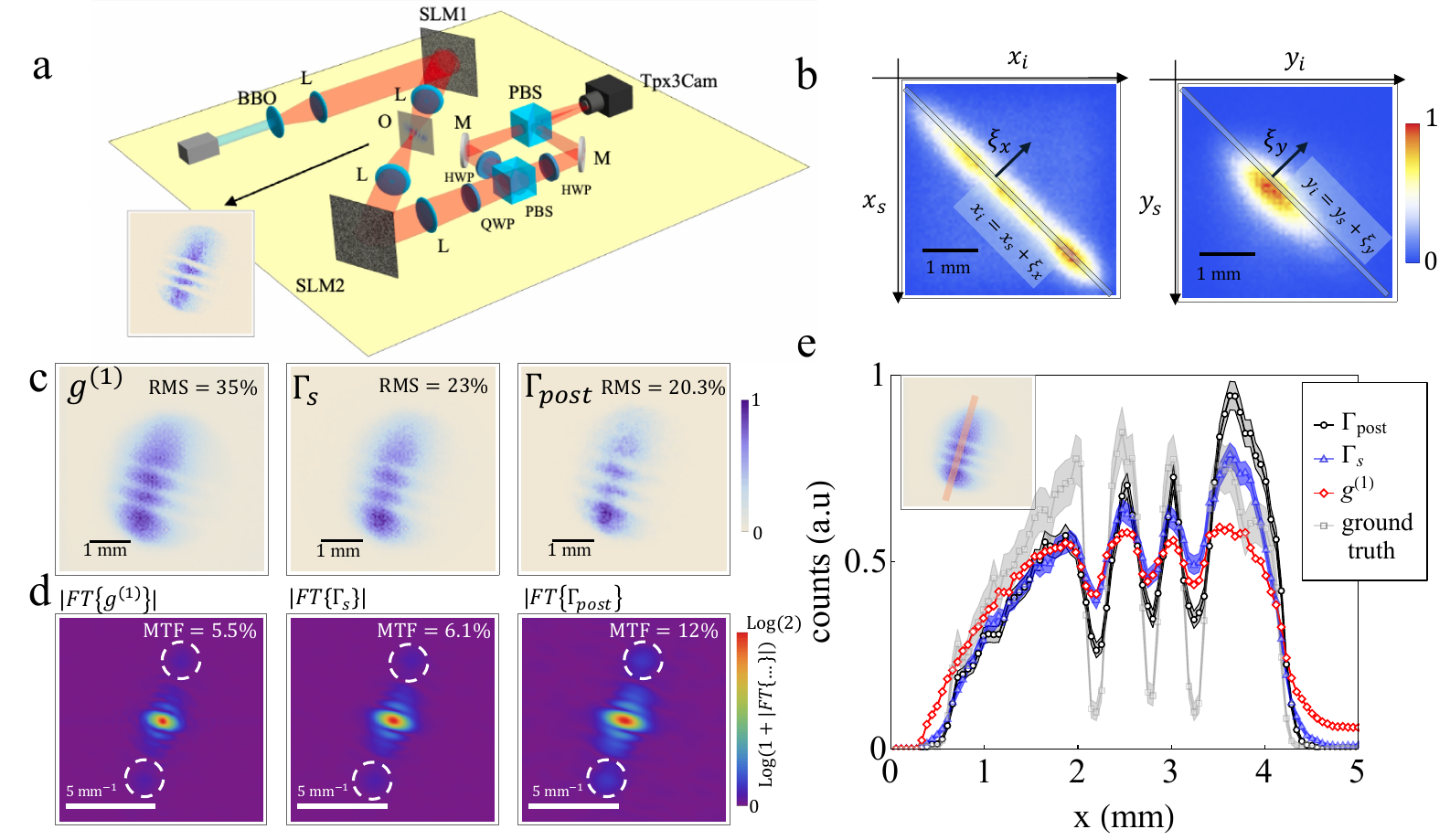}
  \caption{\textbf{Experimental setup and results.} \textbf{a,} Outline of the experimental setup. BBO, $\beta$-barium borate crystal cut for Type I SPDC. SLM1 and SLM2 are liquid crystal spatial light modulators used to introduce scattering in S1 and S2 planes, respectively. O is the amplitude sample to be imaged. L, Lens; M, mirror; BS, beam splitter. Spatially resolved coincidences are collected by a time-stamping camera, Tpx3Cam. \textbf{b,} Near field correlations after scattering. Coincidence postselection is obtained by selecting events where signal and idler coordinates are linearly correlated $\mathbf{x}_i=\mathbf{x}_s+\boldsymbol{\xi}$. \textbf{c,} Image of the object immersed in a dynamic scattering medium as reconstructed by singles ($g^{(1)}$), coincidences ($\Gamma_s$), and post-selected coincidences ($\Gamma_{post}$). $\Gamma_{post}$ is obtained by summing $6\times6$ postselection images --i.e. 36 different choices of $\boldsymbol{\xi}$-- to reduce shot noise (see text for details). $\Gamma_{post}$ demonstrates an enhanced contrast, as highlighted in \textbf{d} and \textbf{e}. RMS indicates the root-mean-square error when comparing each output with the ground truth (for which we used the inset of panel \textbf{a}). \textbf{d}, 2D Fourier spatial spectra of the images in \textbf{b}. The contrast enhancement corresponds to the increased relative brightness of the spatial frequencies evidenced by the dashed white circles. All 2D plots are normalized with respect to the maximum. From the results in \textbf{d}, we estimated the MTF to be $5.5\,\%, 6.1\, \%,$ and $12\,\%$, respectively. \textbf{e,} 1D cross-section of each image (indicated by the dashed red line in the inset) with error bands corresponding to 3 times the standard error. The scattering strength for the data shown in this figure was estimated to be approximately $w_q/w_A\approx 3.4$.}
  \label{fig:biphoton-platform}
\end{figure*}

\section{Experimental Results}
\subsection{Two photon scattering}
To verify the contrast enhancement allowed by coincidence postselection, we performed a couple of proof-of-principle experiments using spatially correlated photon pairs generated via spontaneous parametric down-conversion (SPDC) in a \SI{1}{\milli\meter}-thick Type-I $\beta$-barium borate (BBO) crystal pumped by a pulsed \SI{405}{\nano\meter} laser. In the experiment, frequency-degenerate photons at \SI{810}{\nano\meter} were spectrally filtered using bandpass filters with \SI{10}{\nano\meter} bandwidth. Reflective liquid-crystal spatial light modulators (LC-SLMs) were used to introduce random phase patterns onto the down-converted beams. A conceptual scheme of the experiment is sketched in Fig.~\ref{fig:biphoton-platform}-a, which was designed to reproduce the conditions considered in the numerical simulations (Fig. \ref{fig:sim}-a). The object O consisted of 3 dark lines drawn with a black marker on a transparent plastic fragment (see Supplementary Information). The inset in Fig.~\ref{fig:biphoton-platform}-a shows the appearance of the object on the camera plane in the absence of scattering.
In this first configuration, the photon pairs travel along the same beam path but are spatially anti-correlated in the far field, where each photon interacts with a different region of the scattering screens SLM1 and SLM2. The relay imaging system re-images the crystal plane onto the object location, where the photon pairs overlap.
In order to spatially resolve two photo coincidences, signal and idler photons were probabilistically separated in two different optical paths and detected on corresponding regions of a time-stamping camera (Tpx3Cam) equipped with an image intensifier for single-photon detection \cite{nomerotski2023intensified}. The Tpx3Cam records the arrival time and the spatial coordinates of each event with $\sim7$ ns resolution \cite{vidyapin2023characterisation}. From these data, we can extract a list of coincidences between two separate regions of interest (ROIs) on the camera sensor. Here, a coincidence event is defined by two photons arriving within 10\,ns of each other.  % By defining separate regions of interest (ROIs) on the camera sensor that correspond to the signal and idler detection areas, we can record the arrival positions of both photons. 

The resulting data set consists of a compressed format of the four-dimensional coincidence distribution $\Gamma(\mathbf{x}_i, \mathbf{x}_s)$, which records all spatial correlations between photon pairs. An example of spatial correlations in the near field after scattering is shown in Fig.~\ref{fig:biphoton-platform}-b. The presence of scattering media broadens the spatial correlations along both the $x$ and $y$ coordinates, which, in the absence of scattering, are only $\sim 1.5$ pixel wide. 

To isolate ballistic photons from the scattered background, we select events where idler and signal positions differ by a fixed amount: $\mathbf{x}_i = \mathbf{x}_s + \boldsymbol{\xi}$, as illustrated by the rectangular shades in Fig.~\ref{fig:biphoton-platform}-b. The resulting postselected images $\Gamma_{\text{post}}(\mathbf{x} ; \boldsymbol{\xi}):=\Gamma(\mathbf{x}_i= \mathbf{x}_s+\boldsymbol{\xi})$ are expected to yield contrast enhancement. This spatial filtering preferentially retains ballistic photons that maintain their correlation, while rejecting scattered photons that no longer exhibit spatial correlation. At the same time, due to the lower number of associated events, $\Gamma_{\text{post}}$ will also exhibit higher shot noise. More events can be included by choosing different offsets $\boldsymbol{\xi}$, which give only a rigid translation of the corresponding postselected image.
 
The final images are obtained by summing $n\times n $ postselected images after each image has been appropriately shifted:  $\Gamma^{(n)}_{\text{post}}(\mathbf{x}) = \sum_{\xi_x, \xi_y = 1, \ldots, n} \Gamma_{\text{post}}(\mathbf{x} - \boldsymbol{\xi}, \boldsymbol{\xi})$. For all the experimental acquisitions, we recorded 500 frames at \SI{20}{\second} per frame.  

% After propagating through this scattering environment, a quarter-wave plate and polarizing beam splitter separate the photon pairs onto different camera regions.
%Photon events are identified in post-processing by clustering pixel hits and centroiding each cluster to estimate photon positions \cite{ianzano2020fast}. Coincidence events are then extracted using a \SI{10}{\nano\second} temporal correlation window \cite{edgar2012imaging}. 

Figure~\ref{fig:biphoton-platform}-c shows a comparison between classical single-photon imaging ($g^{(1)}$), non-post-selected coincidence imaging ($\Gamma_s$), and post-selected coincidence imaging ($\Gamma^n_{\text{post}}$, with $n=6$) for a test object immersed in dynamic scattering media. The scattering strength for this measurement was estimated to be approximately $w_q/w_A \approx 5.7$. The estimate was made by measuring the ratio of the measured correlation widths with and without scattering. The non-post-selected coincidence image $\Gamma_s$ shows only marginal improvement when compared with the direct image $g^{(1)}$. In contrast, the post-selected coincidence image $\Gamma^{(n=6)}_{\text{post}}$ reveals the object features with dramatically enhanced contrast. Results for other values of $n$ are shown in the supplementary.
Each image was compared with the ground truth (inset in panel a), by evaluating the root-mean-square difference $\text{RMS}:=\sum_{i}|I^{g}_i-I^{r}_i|/\sum_{h}(I^{g}_h+I^{r}_h)$ with $I^g$ being the ground truth image and $I^r=g^{(1)},\Gamma_s, \Gamma_{post}$. Coincidence images show an improvement in the $RMS$ with respect to the ground truth; however, the relative improvement for $\Gamma_{post}$ with respect to $\Gamma_{s}$ is marginal due to the increased noise in the postselected image. The advantage of the postselection is more evident when targeting a measure of the contrast, which is more robust to noise fluctuations. 
%For this figure, we summed $6 \times 6 = 36$ different post-selected images with varying offset vectors $\boldsymbol{\xi}$ to reduce shot noise while maintaining contrast enhancement. 

The contrast enhancement is quantified through the modulation transfer function (MTF), calculated from the 2D Fourier spatial spectra shown in Fig.~\ref{fig:biphoton-platform}-d. %The Fourier transforms are displayed on a logarithmic color map, $\log(1 + |\mathcal{FT}\{g^{(1)}\}|)$, $\log(1 + |\mathcal{FT}\{\Gamma_s\}|)$, and $\log(1 + |\mathcal{FT}\{\Gamma_{\text{post}}\}|)$, respectively.
All 2D plots are normalized with respect to their respective maxima. From the results in Fig.~\ref{fig:biphoton-platform}-d, we estimated the MTF to be 5.5\% for the singles ($g^{(1)}$), 6.1\% for non-post-selected coincidences ($\Gamma_s$), and 12\% for post-selected coincidences ($\Gamma_{\text{post}}$). The MTF was obtained from the ratio between the maximum values within the regions highlighted by the dashed circles and the central (zero-frequency) maximum. Post-selection provides more than a two-fold improvement in MTF over the direct imaging and nearly doubles the MTF compared to non-post-selected coincidences. %The increased relative brightness of higher spatial frequencies (indicated by the dashed white circles in Fig.~\ref{fig:biphoton-platform}d) demonstrates the post-selection successfully recovers fine image features lost in scattering. 

The contrast enhancement is further illustrated in the 1D cross-sections shown in Fig.~\ref{fig:biphoton-platform}-e. %The spatial profiles of $g^{(1)}$, $\Gamma_s$, and $\Gamma_{\text{post}}$ are the red, blue, and black dashed lines, respectively. 
While the singles profile and the non-post-selected coincidence profile show only faint object features, the post-selected profile exhibits significantly enhanced contrast. The error bands, corresponding to three times the standard error, confirm that the observed differences between profiles are well above the measurement uncertainty. This demonstrates that spatial correlations in biphoton states can effectively filter ballistic photons from scattered background, even in the presence of dynamic scattering. Moreover, the fact that the error band in $\Gamma^{(n=6)}_{pos}$ is of the same order than for the other two datasets shows how summing over multiple postselected images helps reducing the shot noise by retaining the contrast enhancement.

\begin{figure}[!ht]
\includegraphics[width=0.99\columnwidth]{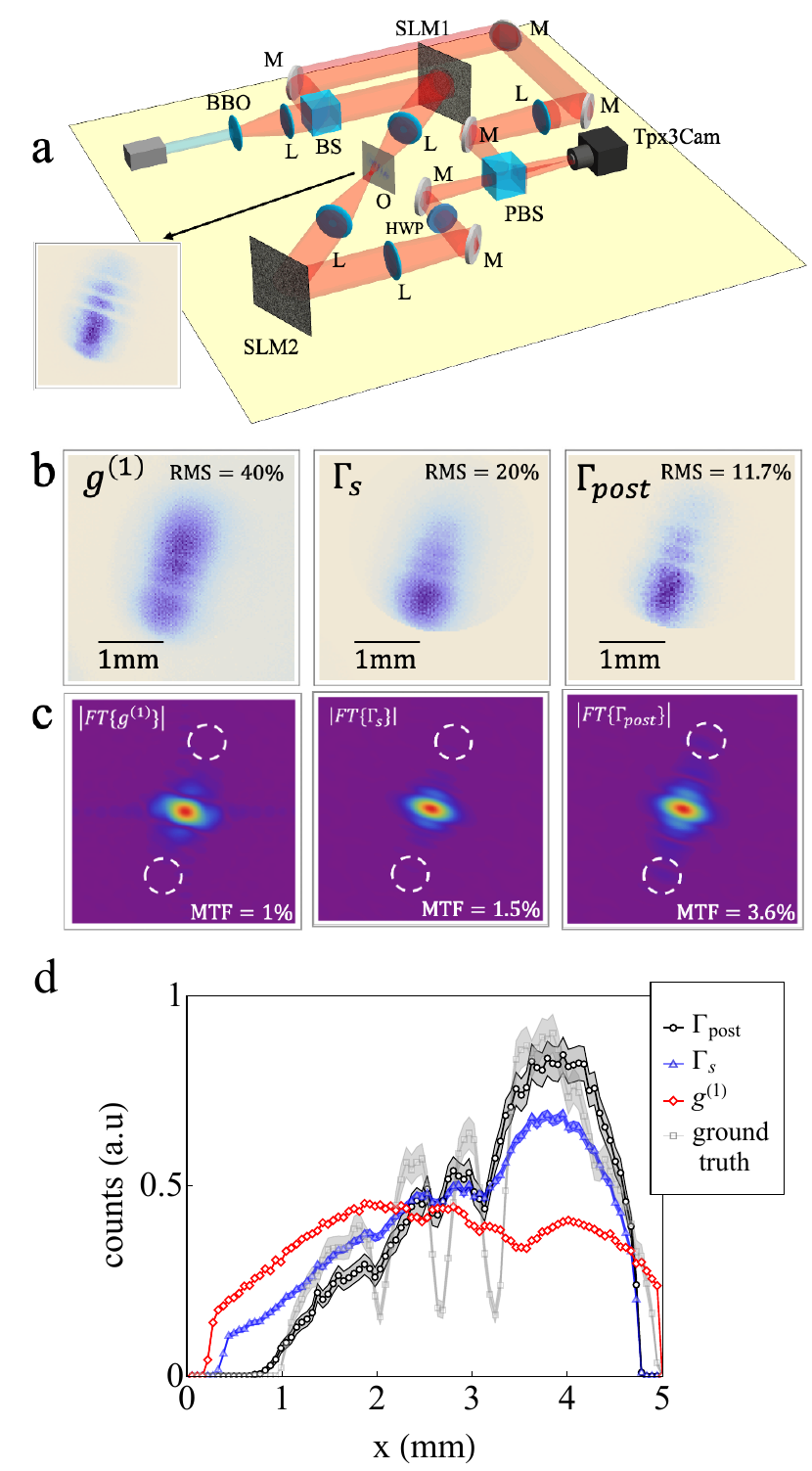}
\caption{\textbf{Correlation imaging with dynamic scattering.} \textbf{a,} Outline of the experimental setup for correlation imaging. BBO, $\beta$-barium borate crystal cut for Type I SPDC. A beam splitter (BS) in the far field separates signal and idler photons into an object arm and a reference arm. SLM1 and SLM2 are liquid crystal spatial light modulators used to introduce scattering. O is the amplitude sample to be imaged. L, Lens; M, mirror; BS, beam splitter. Spatially resolved coincidences are collected by a time-stamping camera, Tpx3Cam. \textbf{b,} Image of the object immersed in a dynamic scattering medium as reconstructed by singles ($g^{(1)}$), coincidences ($\Gamma_s$), and post-selected coincidences ($\Gamma^{(n=3)}_{\text{post}}$). \textbf{c,} 2D Fourier spatial spectra of the images in \textbf{b}. The contrast enhancement corresponds to the increased relative brightness of the spatial frequencies encircled by the dashed white circles. The scattering strength for the data shown in the figure was estimated to be approximately $w_q/w_A \approx$ 6.8. \textbf{c,} 1D cross section with errors corresponding to 3 times the standard deviation.}
\label{fig:corr_imaging}
\end{figure}

\subsection{Single photon scattering}

Until now, we have considered the configuration in which both photons propagate through the scattering medium. In several quantum imaging experiments, including ghost imaging and correlation imaging, only one of the two photons passes through the object and surrounding scattering media. This may often be a preferred scenario, since, for a fixed scattering medium, the spatial correlation broadening is expected to be less severe when only one photon is affected than when both photons traverse the same scattering screens.  
%In this configuration, a D-shaped mirror is placed in the far-field of the crystal, which is used to separate the signal and idler photons into the object arm and reference arm, respectively. 
To demonstrate this approach, we implemented a correlation imaging configuration where signal and idler photons are separated immediately after generation (Fig.~\ref{fig:corr_imaging}a). A beam splitter positioned in the far field of the crystal directs photons into an object arm and a reference arm. In the object arm, the signal photon traverses through scattering screen SLM1, the object positioned at the crystal's image plane, and scattering screen SLM2. The reference arm photon propagates directly to a separate region of the camera sensor without encountering any scattering. Spatially resolved coincidences between the two arms are then recorded with a time-stamping camera (Tpx3Cam). The data acquisition and analysis procedures follow those described for Fig.~\ref{fig:biphoton-platform}, with 500 frames at \SI{20}{\second} per frame. Figure~\ref{fig:corr_imaging}b shows the comparison between singles ($g^{(1)}$), non-post-selected coincidences ($\Gamma_s$), and post-selected coincidences ($\Gamma^{(n=3)}_{\text{post}}$) for an object imaging through dynamic scattering in the object arm. The singles' image shows degradation due to scattering. The non-post-selected coincidence image ($\Gamma_s$) shows only marginal improvement. Instead, the post-selected image ($\Gamma^{(n=3)}_{\text{post}}$ reveals enhanced contrast, demonstrating that the post-selection can recover correlation information even when only one photon interacts with the object.

The contrast enhancement is again evident when analyzing the MTF. The 2D Fourier spatial spectra in Fig.~\ref{fig:corr_imaging}c, demonstrate the preservation of higher spatial frequencies in the post-selected image (indicated by the dashed white circles). From the Fourier spectra, we estimated the MTF to be 4.52\% for singles ($g^{(1)}$), 3.55\% for non-post-selected coincidences ($\Gamma_s$), and 6.92\% for post-selected coincidences ($\Gamma^{(n=3)}_{\text{post}}$). Post-selection provides approximately 50\% improvement over singles and nearly doubles the MTF compared to non-post-selected coincidences. This demonstrates that correlation-based filtering remains effective even when only one photon interacts with the object and scattering medium, while the reference photon preserves the spatial correlation information needed for post-selection.

\begin{figure}[!ht]
\includegraphics[width=0.9\columnwidth]{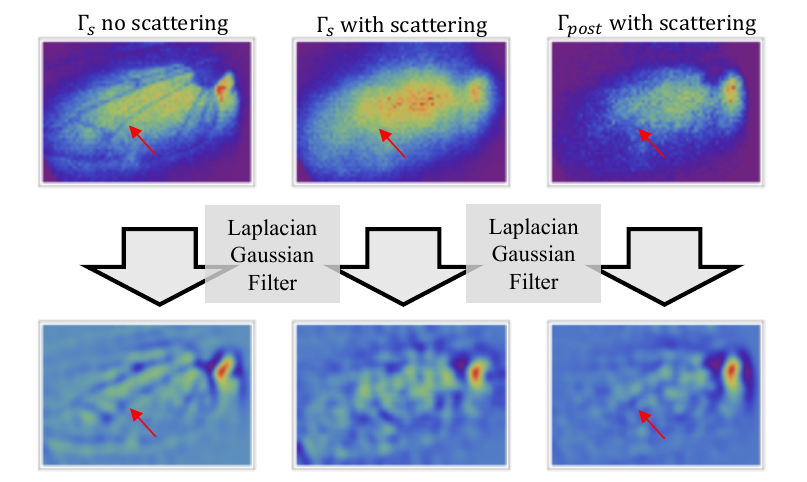}
\caption{{\bf Retrieving image details from correlation postselection.} Postselection of correlation data allows to recover object details that would otherwise be completely scrambled by the scattering medium. In this example, where the object is an insect wing, it is seen how scattering hides the venation pattern. Some of the veins are, however, slightly visible when looking at the postselection data. This is more evident when digital filtering is applied to the recovered images. The corresponding images in the absence of scattering (before and after digital filtering) are shown on the left for comparison. These data were acquired in the single-photon-scattering configuration. }
\label{fig:wing}
\end{figure}

As an applied example, we illustrate how the postselected images can allow for identifying features otherwise not recognizable by direct or coincidence imaging. In the example reported in Fig. \ref{fig:wing}, the object consisted of an insect wing. Under the effect of sufficiently strong scattering-- here applied on one of the two photons--fine details, like veins, become completely unrecognizable, even when processing the image with digital filters. As shown in Fig. \ref{fig:wing}, postselection images can instead help to identify these details, as is more evident when a Laplacian-Gaussian filter is applied.

\section{Conclusion}
We have demonstrated that spatial post-selection of correlated photon pairs enables contrast-enhanced imaging through dynamic and static scattering media. Our experimental results show that post-selection can recover object features that are generally obscured by scattering, with modulation transfer function improvements of up to two-fold over non-post-selected coincidences and several-fold over classical single-photon imaging. This enhancement persists across different experimental configurations, including biphoton imaging, where both photons traverse scattering screens and correlation imaging, where only one photon interacts with the object. 

The key mechanism underlying this enhancement is the spatial filtering of ballistic photons through coincidence post-selection. By selecting photon pairs that maintain specific spatial correlations $\mathbf{x}_i = \mathbf{x}_s + \boldsymbol{\xi}$, we preferentially retain photons that have not been significantly scattered, while rejecting those that have lost their correlation due to random phase shifts. 

Future work could explore several promising directions for improvement and extension. While here we focused on the detection of amplitude objects, possible generalizations to phase objects \cite{zia2023interferometric,ortolano2023quantum,dehghan2024biphoton, dehghan2026quantum} remain to be tested. 

The approach introduced in this work could offer advantages over other classical techniques in experiments involving thin scattering media, where scattered photons experience strong phase distortions but small -- below ps -- temporal delays, making time gating difficult. Moreover, this technique is directly applicable in applications requiring low illumination to avoid photo-damage. Further suppression of scattered photons could be achieved by combining this technique with temporal gating employing the pump laser for a nonlinear optical switch \cite{xu2026epsilon, duguay1969ultrafast}. 
Furthermore, integrating adaptive optics strategies could be employed to mitigate higher levels of scattering \cite{cameron2024adaptive, scarfe2025fast, verniere2026entanglement,courme2026non}, hypothetically enabling imaging through even more strongly scattering media. 
\section{Acknowledgments}
This work was supported by the Canada Research Chair (CRC) Program, NRC-uOttawa Joint Centre for Extreme Quantum Photonics (JCEP) via the Quantum Sensors Challenge Program at the National Research Council of Canada, and Quantum Enhanced Sensing and Imaging (QuEnSI) Alliance Consortia Quantum grant.

\bibliography{bibliography.bib}

\vspace{1 EM}

\noindent\textbf{Author Contributions}\noindent
AD and YK conceived the idea. AD developed the simulations. JH, RA, ND, and AD, with contributions from YK and YZ, built the experimental setup and collected the data. YZ, with contributions from AD, prepared the software for the extraction of post-selected correlation images. AD, YK, JH, and RA analyzed the results. AD, JH, and RA prepared the first version of the manuscript. EK supervised the project.

\vspace{1 EM}

\noindent\textbf{Data availability}
\noindent
The data that support the findings of this study are available from the corresponding author upon reasonable request.
\vspace{1 EM}

\noindent\textbf{Code availability}
\noindent
The code used for the data analysis is available from the corresponding author upon reasonable request.

\vspace{1 EM}
\noindent\textbf{Ethics declarations} Competing Interests. The authors declare no competing interests.

\vspace{1 EM}
\noindent\textbf{Corresponding authors}
Correspondence and requests for materials should be addressed to aderrico@uottawa.ca.

\clearpage
\onecolumngrid
\renewcommand{\figurename}{\textbf{Figure}}
\setcounter{figure}{0} \renewcommand{\thefigure}{\textbf{S{\arabic{figure}}}}
\setcounter{table}{0} \renewcommand{\thetable}{S\arabic{table}}
\setcounter{section}{0} \renewcommand{\thesection}{S\arabic{section}}
\setcounter{equation}{0} \renewcommand{\theequation}{S\arabic{equation}}
\onecolumngrid
\begin{center}
{\Large Supplementary Material for: \\Contrast enhanced imaging through weakly scattering media with spatially entangled photons.}
\end{center}
%\appendix
\vspace{1 EM}
%%%%%%%%%%%%%%%%%% SM %%%%%%%%%%%%%%%%%%%%%%%

\section{Generation and characterization of scattering strength}

The measured spatial correlations can be used to estimate the scattering strength $w_q/w_A$. In the case of the biphoton imaging, we find that a good estimate can be obtained from the relation (derived semiempirically from numerical simulations --see figure S4--) 
\begin{align}
    w_-/w_{-,0}=\sqrt{1+3(w_q/w_A)^2}
\end{align}
where $w_{-,0}=2/w_q$ is the correlation width without scattering, and $w_{-}$ the correlation width with scattering. In the case of correlation imaging the equation must be corrected as  \begin{align}
    w_-/w_{-,0}=\sqrt{(1+3(w_q/w_A)^2)/2}.
\end{align}
The obtained results are in qualitative agreement with a direct measurement where the phase-matching intensity on the SLM plane was measured and directly compared with the applied holograms. However, the phase-matching function exhibits a longitudinal walk-off for which the Gaussian approximation is only a rough estimate.

\section{Result for different n scans}

Figure S3 illustrates the effect of combining multiple post-selected images to improve the signal-to-noise ratio. It is seen that, while signal-to-noise ratio improves by increasing $n$, there is also a decrease in contrast, indicating the existence of a tradeoff in choosing the optimal $n$. The contrast of different experimental results is shown in Table S1.
\begin{table}[h!]
\centering
\begin{tabular}{|c|c|c|c|c|c|}
\hline
 & $\Gamma^{1}_{post}$(\%) & $\Gamma^{3}_{post}$(\%) & $\Gamma^{6}_{post}$(\%) &  $\Gamma_{s}$(\%) & $g^{(1)}$(\%) \\
\hline
biphoton imaging ( Fig 4) & 21.7 & 15.6 & 9.5 & 5.3 & 3.0 \\
\hline
biphoton imaging (Fig. 3) & 12.3& 8.4 &6.2  & 4.3 &  2.6\\
\hline
correlation imaging / low scattering (Fig. 5) &10.8  &9.0&9.0  & 4.8 &3.4\\
\hline
correlation imaging / high scattering & 4.0 & 3.8 & 3  &  1.1 & 0.9 \\
\hline
 
\end{tabular}
\caption{\bf{MTF values for different $\Gamma^{(n)}_{post}$ compared to the coincidences $\Gamma_{s}$ and singles $g^{(1)}$}}.
\end{table}

\section{Experimental results for static scattering.}
To further validate the approach under different experimental conditions, we considered a configuration with a single static random mask --applied on SLM2--,  placed between the object and the detector. Without scattering before the object, strong spatial correlations are preserved on the object plane. In this configuration, in the absence of any scattering, coincidence imaging is known to give a two-fold enhanced contrast. This enhancement is still observed when SLM2 applies a fixed random phase pattern. As in the previous case, a further enhancement is obtained when considering postselected images, as evidenced by the Fourier spatial spectra \ref{fig:3slit SCANS}-b, and the cross sections \ref{fig:3slit SCANS}-c. 

From the Fourier spectra, we estimated the MTF to be 3.0\% for singles ($g^{(1)}$), 5.3\% for non-post-selected coincidences ($\Gamma_s$), and 15.6\% for post-selected coincidences $\Gamma^{(n=3)}_{pos}$. Post-selection provides more than a five-fold improvement over singles and approximately a three-fold improvement compared to the non-post-selected coincidences.% This clearly demonstrates that correlation-based filtering remains effective when photons interact with the object before a scattering environment, preserving the spatial relationship needed for post-selection. 

%MTF for $\Gamma^1_{\text{post}}$ is 21.7\%
%MTF for $\Gamma^3_{\text{post}}$ is 15.6%
%MTF for $\Gamma^6_{\text{post}}$ is 9.5\%
%MTF for $\Gamma_s$ is 5.3\%
%MTF for singles ($g^{(1)}$) is 3.0\%

\clearpage
\begin{figure*}
\includegraphics[width=0.8\columnwidth]{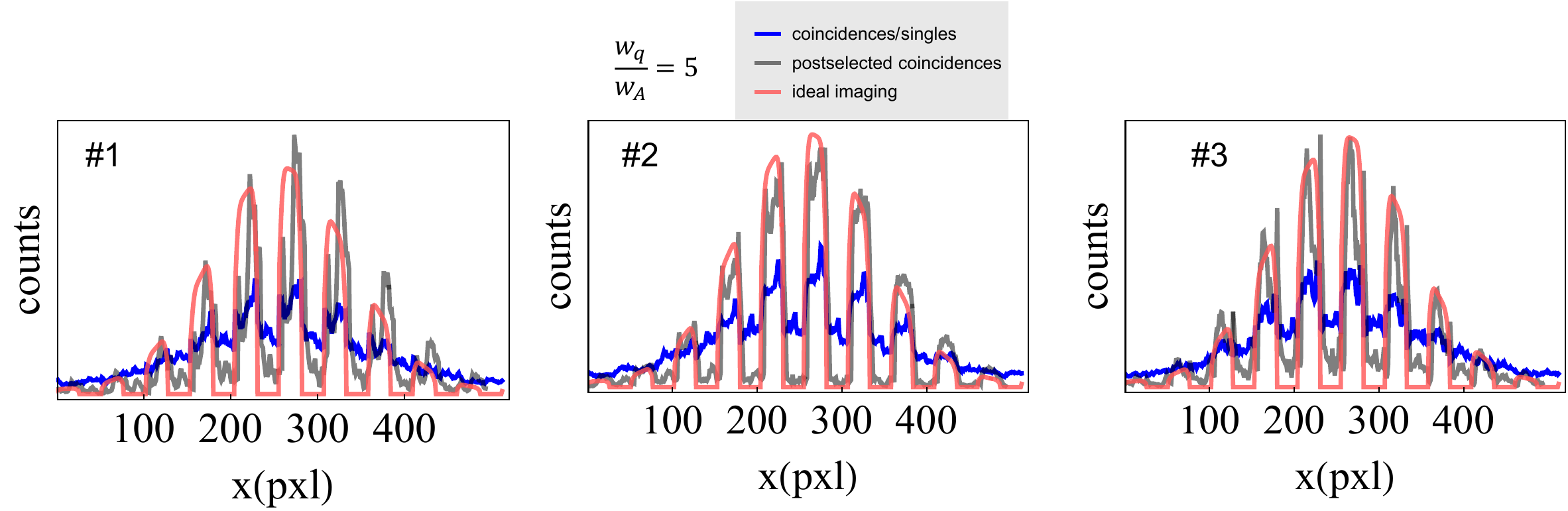}
\caption{{\bf Simulated results for single realizations of scattering}. Examples of 3 different realizations of the effect of scattering on the MTF. In all cases, postselection leads to contrast improvement, although image distortions, appearing as count fluctuations, are introduced. } 
\label{fig:staticsim}
\end{figure*}
\begin{figure*}
\includegraphics[width=0.8\columnwidth]{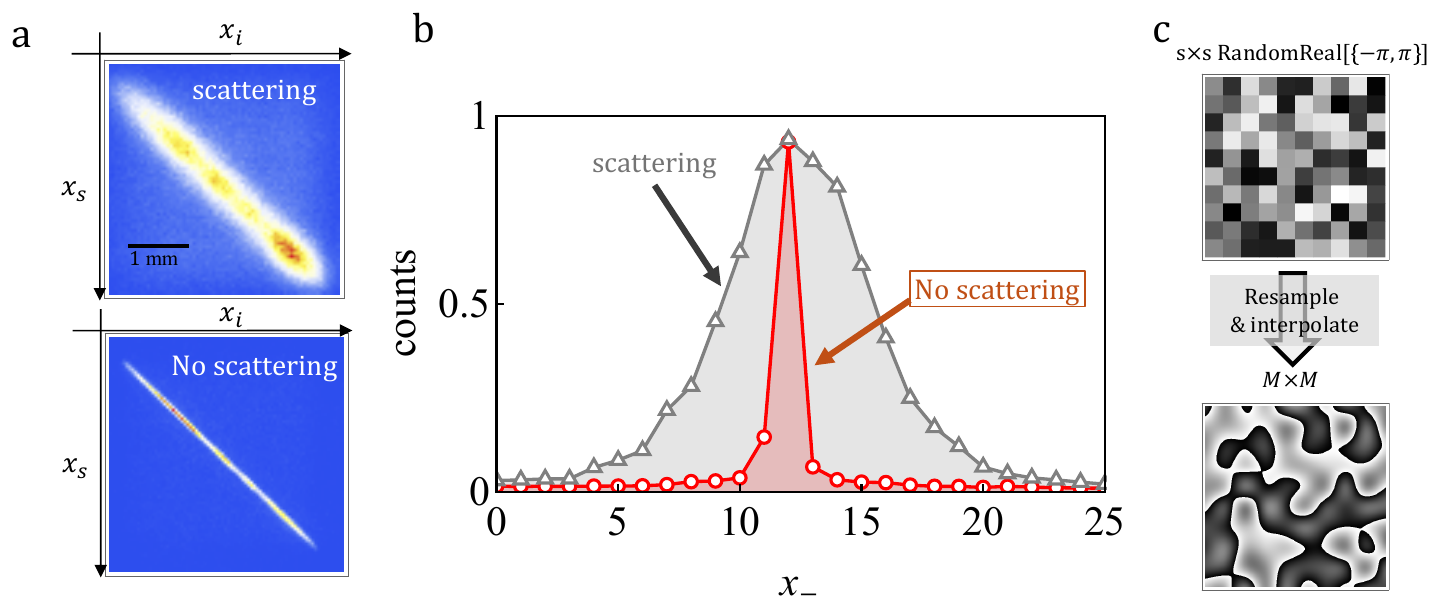}
\caption{{\bf Scattering characterization and SLM mask generation}. (a) Measured correlations with and without scattering with correlation broadening highlighted in (b). (c) Illustration of the procedure for preparing the phase masks. } 
\label{fig:scattcharact}
\end{figure*}

\begin{figure*}
\includegraphics[width=0.8\columnwidth]{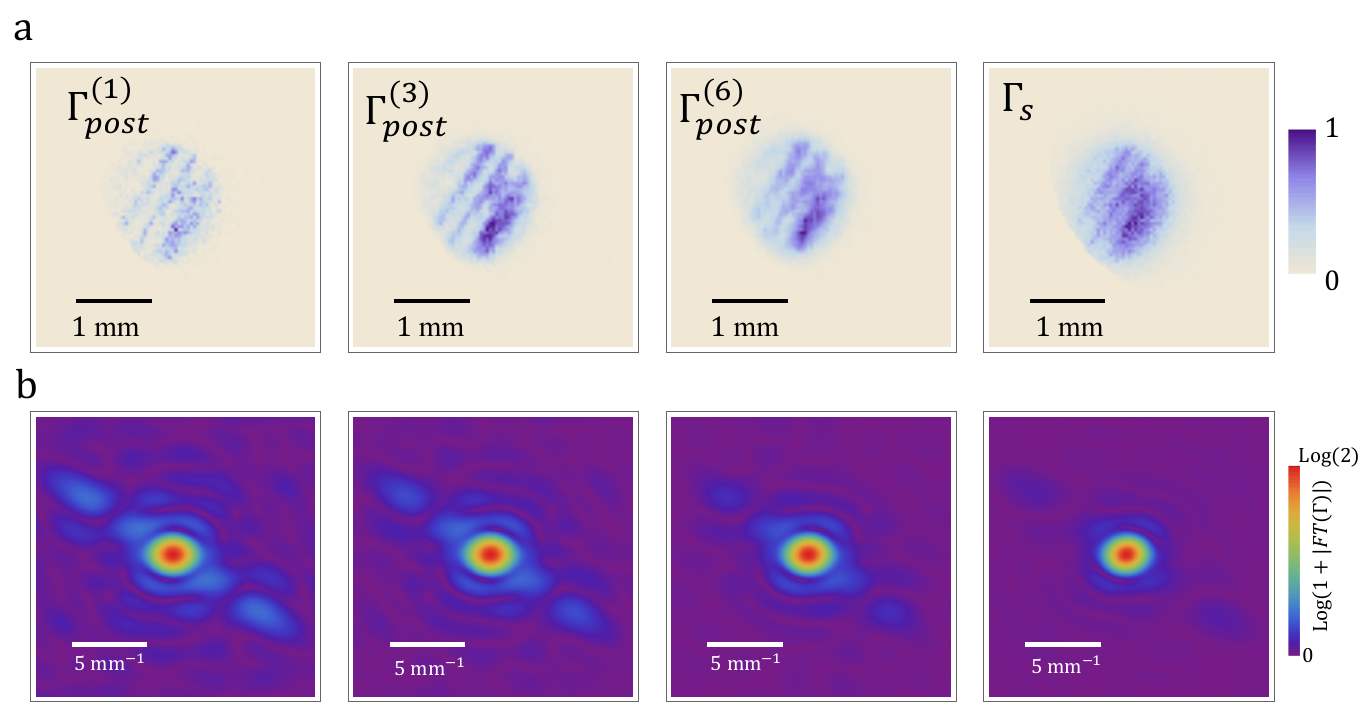}
\caption{{\bf Results for different numbers of postselection $n$.} } 
\label{fig:manyscans}
\end{figure*}

\begin{figure*}
\includegraphics[width=0.8\columnwidth]{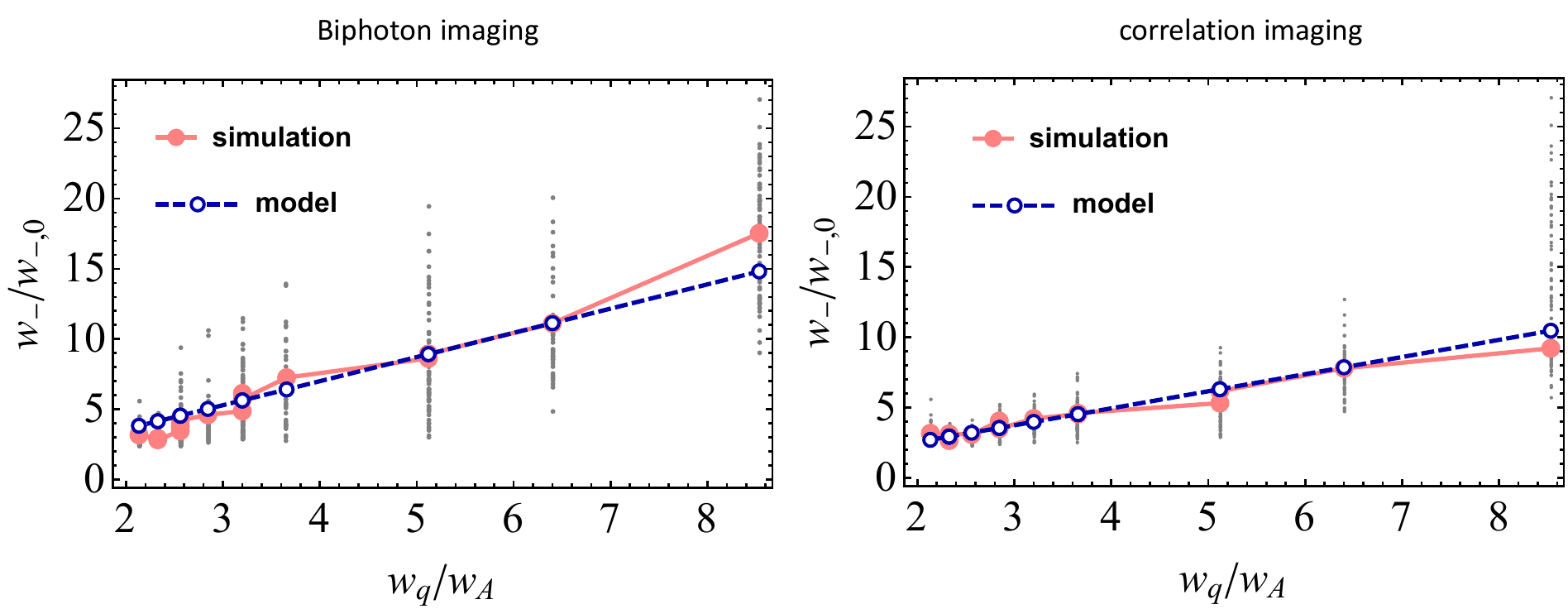}
\caption{{\bf Simulated correlation broadening as a function of correlation strength }. Simulations were carried out considering 40 realizations (gray dots) of the same scattering strength and averaging over the results. The simulation window had $M=2^7$ pixels, with $w_q=0.2 M$ and $w_{0}=w_q/15$. The results are compared with formulas S1 and S2 for the biphoton and correlation imaging, respectively.} 
\label{fig:scattcharact}
\end{figure*}

\begin{figure*}[t]
\includegraphics[width=\textwidth]{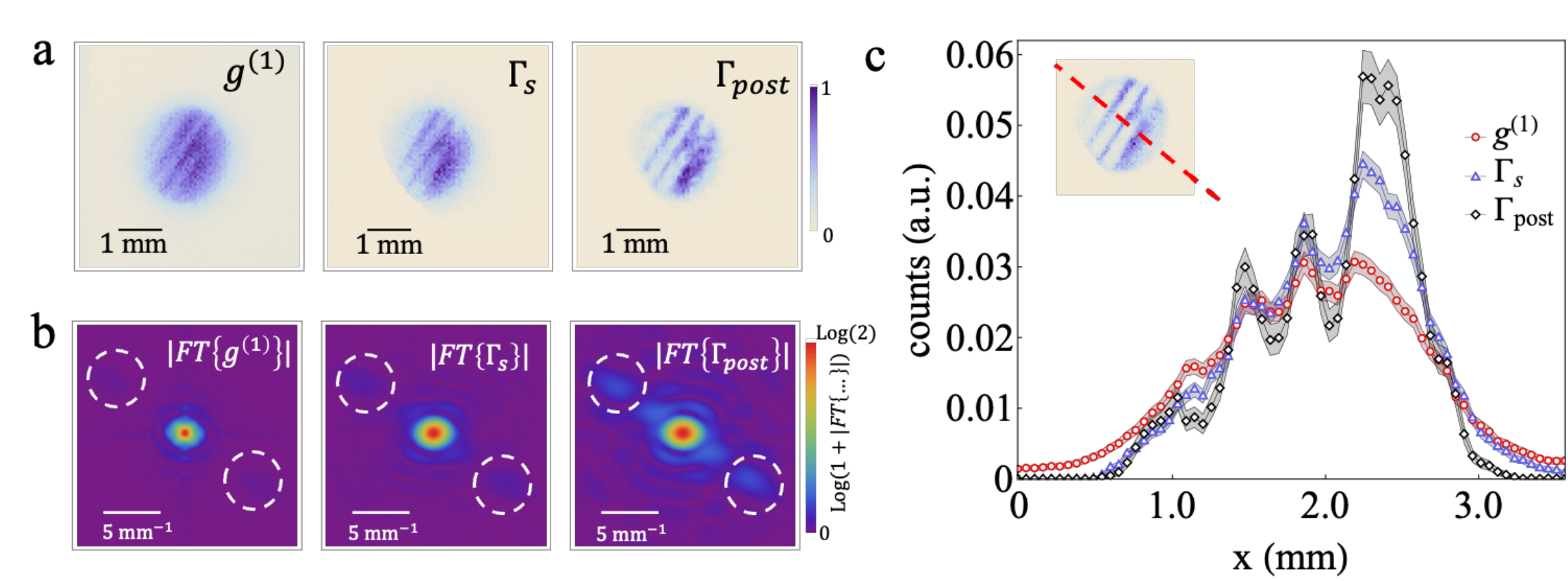}
\caption{ {\bf Contrast enhancement with scattering medium between camera and object.} \textbf{a} Image of the object hidden behind a static scattering medium as reconstructed by singles ($g^{(1)}$), coincidences ($\Gamma_s$), and post-selected coincidences ($\Gamma^{(3)}_{post}$). $\Gamma^{(3)}_{post}$ demonstrates an enhanced contrast, as highlighted in \textbf{b} and \textbf{c}. \textbf{b}, 2D Fourier spatial spectra of the images in \textbf{a}. The contrast enhancement corresponds to the increased relative brightness of the spatial frequencies evidenced by the dashed white circles. All 2D plots are normalized with respect to the maximum. From the results in \textbf{b}, we estimated the MTF to be $3.05\,\%, 5.3\, \%,$ and $15.6\,\%$, respectively. \textbf{c,} 1D cross-section of each image (indicated by the dashed red line in the inset) with error bands corresponding to 3 times the standard error. In this example, a contrast enhancement is also seen for $\Gamma_s$ due to the two photons being correlated on the object plane. The scattering strength is, as in Fig. \ref{fig:biphoton-platform}, $w_q/w_A\approx 3.4$. %[n=3]. MTF=3.05, 5.26, 15.59
}
\label{fig:3slit SCANS}
\end{figure*}
\end{document}